\documentclass[sigconf]{acmart}

\usepackage{booktabs} % For formal tables
\usepackage{threeparttable}
\usepackage[official]{eurosym}
\usepackage{todonotes}
\usepackage{comment}
\usepackage{multirow}
\usepackage{float}

\usepackage{amssymb}% http://ctan.org/pkg/amssymb
\usepackage{pifont}% http://ctan.org/pkg/pifont
\newcommand{\cmark}{\ding{51}}%
\newcommand{\xmark}{\ding{55}}%

\setcopyright{rdsm2018}

\acmConference[RDSM]{Workshop on Rumours and Deception in Social Media}{October 2018}{Turin, Italy}
\begin{document}
\title{FTR-18: Collecting rumours on football transfer news}
%\titlenote{Produces the permission block, and  copyright information}
%\subtitle{Extended Abstract}
%\subtitlenote{The full version of the author's guide is available as   \texttt{acmart.pdf} document}

\author{Danielle Caled,   M\'ario J. Silva}
\affiliation{
  \institution{INESC-ID Lisboa, Instituto Superior T\'ecnico, Universidade de Lisboa}
  \streetaddress{Rua Alves Redol 9}
  \state{Lisbon, Portugal}
}
\email{dcaled@inesc-id.pt, mjs@inesc-id.pt}

\begin{abstract}

This paper describes ongoing work on the creation of a multilingual rumour dataset on football transfer news, FTR-18. Transfer rumours are continuously published by sports media. They can both harm the image of player or a club or increase the player's market value. The proposed dataset includes transfer articles written in English, Spanish and Portuguese. It also comprises Twitter reactions related to the transfer rumours. FTR-18 is suited for rumour classification tasks and allows the research on the linguistic patterns used in sports journalism.

\end{abstract}

%
% The code below should be generated by the tool at
% http://dl.acm.org/ccs.cfm
% Please copy and paste the code instead of the example below.
%

\begin{CCSXML}
<ccs2012>
<concept>
<concept_id>10010147.10010178.10010179</concept_id>
<concept_desc>Computing methodologies~Natural language processing</concept_desc>
<concept_significance>500</concept_significance>
</concept>
</ccs2012>
<ccs2012>
<concept>
<concept_id>10002951.10003227.10003233</concept_id>
<concept_desc>Information systems~Collaborative and social computing systems and tools</concept_desc>
<concept_significance>500</concept_significance>
</concept>
<concept>
<concept_id>10010147.10010178.10010179</concept_id>
<concept_desc>Computing methodologies~Natural language processing</concept_desc>
<concept_significance>500</concept_significance>
</concept>
</ccs2012>
\end{CCSXML}

\ccsdesc[500]{Information systems~Collaborative and social computing systems and tools}
\ccsdesc[500]{Computing methodologies~Natural language processing}

\keywords{football transfer rumours, rumour detection, rumour classification}

\maketitle

\section{Introduction}

In the last years, the way news are spread has changed. Social networks, blogs/micro-blogs and other untrusted online news sources have gained popularity, thus allowing any user to produce unverified content. This modern kind of media enables real-time proliferation of news stories, and, as consequence, increases the diffusion of rumours, hoaxes and misinformation to a global audience \cite{tacchini2017some}. As news content is continuously published online, the speed in which it is disseminated hinders human fact-checking activity. A piece of information whose ``veracity status is yet to be verified at the time of posting'' is called rumour \cite{zubiaga2018detection}. News articles, scientific researches and conspiracy-theory stories can float in the veracity spectrum. They are characterised by distinct stylistic dimensions and these features enable their spread, but are orthogonal to their truthfulness \cite{vosoughi2018spread}. 

The football transfer market offers a fertile ground for rumour dissemination because rumours and misinformation spreads may be motivated by personal profit or public harm. Releasing false or misleading news about alleged moves of players between clubs emerges as a strategy to increase a player's market value and transfer fees. Negligent transfer announcements are assigned to high sums directly related to some transfers \cite{maia2016jornalismo}. In addition, news organisations also benefit from announcements of alleged transfer moves, attracting public attention by selling contents and advertisements. Maia showed that most of the football transfer news published by the three biggest Portuguese sports diaries during the Summer of 2015 were not confirmed \cite{maia2016jornalismo}. Table \ref{tab:mbappe_rumour} provides an example of a news article about a transfer to Real Madrid that was latter denied by the club in an official announcement. Both the transfer news and the announcement were released on July 4, 2018.

According to the Union of European Football Associations' (UEFA) annual report\footnote{https://goo.gl/FoAv2g}, the European transfer market reached a record revenue of \texteuro5.6 billion during the Summer of 2017, a 6\% increase in spending during this period relatively to the highest value recorded in the last ten years. Economical side effects of rumours on football clubs shares are also reported, like the rumour of Cristiano Ronaldo joining Juventus FC, which made the club shares jump almost 10\% on a single day\footnote{https://goo.gl/3wKhj3} before any official announcement.

\begin{table}
  \caption{Example of a transfer rumour.}
  \label{tab:mbappe_rumour}
    \begin{threeparttable}
    \begin{tabular}{p{0.95\columnwidth}}
    \toprule
    \textbf{Topic}: Real Madrid agreed a deal for PSG's Kylian Mbapp\'e. \\
    
    \midrule
    \textbf{Agreeing news source}: www.msn.com\tnote{a} \\
    \textbf{Headline}: Real Madrid agree stunning \texteuro272m deal to sign Kylian Mbappe \\
    \textbf{Extract}: \textit{Real Madrid have reportedly agreed an eye-catching \texteuro272 million deal with Paris Saint-Germain for Kylian Mbappe.} \\

    \textit{According to reports coming out of France from Baptiste Ripart, the French champions are set to lose Mbappe, with Madrid agreeing to pay the fee over four instalments.} \\
    \midrule
    \textbf{Evidence}: Real Madrid official announcement\tnote{b}\\
    \textbf{Extract}: \textit{Given the information published in the last few hours regarding an alleged agreement between Real Madrid C.F. and PSG for the player Kylian Mbapp\'e, Real Madrid would like to state that it is completely false.} \\
 
    \textit{Real Madrid has not made any offer to PSG or the player and condemns the spreading of this type of information that has not been proven by the parties concerned.}\\
    \bottomrule
    \end{tabular}    
    \begin{tablenotes}\footnotesize
    \item [a] https://goo.gl/HhRafF
    \item [b] https://goo.gl/7oCqBy
    \end{tablenotes}
    \end{threeparttable}
\end{table}

Motivated by the extensive attention given by sports media to transfer news and the high amounts involved, we propose the creation of Football Transfer Rumours 2018 (FTR-18), a transfer rumours dataset for researching the linguistic patterns in their text and propagation mechanisms. FTR-18 is designed as a multilingual collection of articles published by the relevant news organisations either in English, Spanish or Portuguese languages. Besides news articles, our proposed dataset also comprises Twitter posts associated to the transfer rumours. The present work describes the creation process of FTR-18 dataset and discusses the intended usage of the collection.

%The novel contribution of this paper is the creation of a multilingual journalistic collection comprising football transfer rumours and Twitter posts discussing those rumours.

%https://www.caughtoffside.com/2018/07/01/real-madrid-willing-to-offer-one-of-two-los-blancos-superstars-plus-e50m-in-attempt-to-land-summer-transfer-for-world-cup-superstar/

%Assim, para Sobral e Magalhães,“as transferências revelam-se uma das principais matérias de notícia para os jornalistas desportivos que, curiosamente, raramente presenciam a negociação propriamente dita” (Sobral & Magalhães, 1999). 

%Além disso, tal como referem os autores , “o facto de se noticiar que o clube x está interessado no jogador y pode fazer subir o valor do passe deste último” (Sobral e Magalhães, 1999:57). 

%Tal como refere Rui Novais, “o jornalismo desportivo oscila entre o imediatismo da oferta informativa ou cobertura noticiosa pura” (Rui Novais, 2010: 13 ). 

\section{Related Work}

Collections of rumours from different natures have been assembled before. However, these datasets mainly focused on political and social issues \cite{qazvinian2011rumor,silverman2015lies,wang2017liar}. Some of these datasets were built using rumours collected from social media platforms \cite{qazvinian2011rumor,zubiaga2016analysing,zubiaga2016learning}, such as Twitter, or with data extracted from fact-checking websites \cite{silverman2015lies,wang2017liar}, like PolitiFact\footnote{http://www.politifact.com/} and Snopes\footnote{https://www.snopes.com/}, while others were crafted using manually created claims based on Wikipedia articles \cite{thorne2018fever}.

The first large-scale dataset on rumour tracking and classification was proposed by \citet{qazvinian2011rumor}. This dataset comprises manually annotated Twitter posts (tweets) from five political and social controversial topics, with tweets marked as \textit{related} or \textit{unrelated} to the rumour. It also provides annotations about users' beliefs towards rumours, i.e., users who endorse versus users who refute or question a given rumour. Twitter posts were also used in the construction of two datasets under the PHEME project \cite{derczynski2014pheme}: PHEME dataset of rumours and non-rumours (PHEME-RNR) and PHEME rumour scheme dataset (PHEME-RSD).
These two collections are composed of tweets from rumourous conversations associated with newsworthy events mainly about crisis situations. PHEME-RNR contains stories manually annotated as \textit{rumour} or \textit{non-rumour}, hence being suited for the rumour detection task \cite{zubiaga2016learning}. On the other hand, PHEME-RSD was annotated for stance and veracity, following crowd-sourcing guidelines \cite{zubiaga2015crowdsourcing}, and tracked three dimensions of interaction expressed by users: support, certainty and evidentiality \cite{zubiaga2016analysing}. PHEME-RSD contains conversations threads in English and in German, however this data is extremely imbalanced as less than 6\% of the tweets are written in German.

% Twitter posts were also used in the construction of two datasets under the PHEME project \cite{derczynski2014pheme}. PHEME-RNR dataset of rumours and non-rumours is a collection of posts and the respective reactions obtained during crisis events \cite{zubiaga2016learning}. This dataset contains stories manually annotated as \textit{rumour} or \textit{non-rumour}, hence being suited for the rumour detection task. The second collection, PHEME stance dataset, is composed of tweets from rumourous conversations associated with nine newsworthy events mainly about crisis situations \cite{zubiaga2016analysing}. Data annotation was performed through crowd-sourcing guidelines \cite{zubiaga2015crowdsourcing} and tracked three dimensions of interaction expressed by users: support, certainty and evidentiality.

Silverman developed a dataset for the analysis of how online media handles rumours and unverified information \cite{silverman2015lies}. The resulting Emergent database is a collection of online rumours comprising topics about war conflicts, politics and business/technology. Emergent data was generated with the help of automated tools to identify and capture unconfirmed reports early in their life-cycle. Then, rumours were classified according to their headline and body text stances and monitored with respect to social network shares (Twitter, Facebook and Google Plus) and version changes. In a posterior work, Ferreira and Vlachos leveraged Emergent into a dataset focusing veracity estimation and stance classification \cite{ferreira2016emergent}. This modified dataset consists of claims and associated news articles headlines. Claims were labelled with respect to their veracity, while the news articles headlines were categorised according to their stances towards the claim: \textit{supporting}, \textit{denying} or \textit{observing}, if the article does not make assessments about the veracity the claim. 

LIAR is another dataset that could be employed in fact-checking and stance classification tasks \cite{wang2017liar}. LIAR includes manually labelled short statements from PolitiFact. All the statements were obtained from a political context, either extracted from debates, campaign speeches, social media posts, news releases or interviews. Each statement was evaluated for its truthfulness and received a veracity label accompanied by the corresponding evidences. 

Thorne et al. developed FEVER (Fact Extraction and VERification), a large-scale manually annotated dataset focusing on verification of claims against textual sources \cite{thorne2018fever}. FEVER is composed of claims generated from modifications in sentences extracted from introductory sections of Wikipedia pages. The claims were manually classified as \textit{supported}, \textit{refuted} or \textit{not enough info} (if no information in Wikipedia can support or refute the claim). Additional Wikipedia justification evidences concerning supporting and refuting sentences were also recorded. 

The PHEME-RSD and FEVER datasets were employed in stance detection and veracity prediction shared tasks. The PHEME-RSD was employed in the Semantic Evaluation (SemEval) 2017 competition, Task 8 RumourEval\footnote{http://alt.qcri.org/semeval2017/task8/} \cite{derczynski2017semeval}. In the first sub-task (Sub-task A: Stance Classification), participants were asked to analyse how social media users reacted to rumourous stories, while for the second sub-task (Sub-task B: Veracity prediction), participants were asked to predict the veracity of a given rumour. In the FEVER shared task\footnote{http://fever.ai/task.html}, participants should build a system to extract textual evidences either supporting or refuting the claim. 

Despite the variety of existing datasets for rumour analysis, none of them addresses football rumours. Besides, rumours in multilingual environments like the European Football market are yet to be explored by the academic research community. For our purpose, however, it is extremely important to track how news about transfer rumours are covered by sports media in different languages, as UEFA transfers generally occur between distinct European countries. Hence, our proposal differs from the previous work as it introduces a new dataset comprising football transfer rumours in three different languages: English, Spanish and Portuguese.

\section{Dataset building}

The data included in FTR-18 was harvested during the 2018 Summer Transfer Window (STW18). Transfer windows are periods during the year in which football clubs can make international transfers. Every league has the right to choose two annual transfer windows, with the opening and closing dates being defined by the football associations of the league.

The collected material includes both transfer news and rumours and the corresponding reactions to these rumours on Twitter. As we develop a multilingual dataset containing news and comments in English, Spanish and Portuguese involving UEFA clubs, the harvesting is performed during a common period including the transfer windows of England, Spain and Portugal. Accordingly, data is being collected from June 24 until August 31, 2018.

FTR-18 is built in three stages. First, we browsed the media and social networks for an initial assessment of transfer rumours involving top UEFA clubs. Next, we performed both a semi-automated news articles harvesting and a Twitter crawling, selecting news and tweets related to the rumours, respectively. Finally, as transfer moves are confirmed (or not), we annotate rumour veracity.

\subsection{Football clubs selection}

The first step in the creation of the FTR-18 is the selection of the clubs to follow during the Summer Transfer Window of 2018. We picked English, Spanish and Portuguese clubs that played the group stage in 2017-2018 UEFA Champions League. This decision was made based on the languages we are more familiar with. With this restriction, we decided to monitor the following clubs:

\begin{description}
    \item \textbf{England}: Chelsea, Liverpool, Manchester City, Manchester United, Tottenham Hotspur
    \item \textbf{Spain}: Atl\'etico Madrid, Barcelona, Real Madrid, Sevilla
    \item \textbf{Portugal}: Benfica, Porto, Sporting CP
\end{description}

\subsection{Transfer rumours selection}

The next stage of the FTR-18 generation is the collection of transfer news. We selected transfer rumours involving the monitored clubs. The rumours included narratives of permanence or transfer moves concerning football players, that is, whether players are leaving or being hired by the monitored clubs. For consistency reasons, we restricted the rumour selection to players affiliated to clubs from countries having English, Spanish or Portuguese as their majority language. The rumours were manually tracked and extracted from many sports news sources.

For each rumour, we annotated the \textit{target} player, the target's club by the opening of the 2018 Summer Transfer Window (denoted as \textit{source}), and the rumoured \textit{destination} club (See Figure \ref{fig:rumour_attr}). We considered rumours either about the transference of a target to another club (source $\neq$ destination) or about the permanence of the player in the current club (source $=$ destination). 

\subsection{News articles harvesting}

News harvesting involved searching and collecting news articles, published by different news organisations, related to the identified rumours in the languages of the dataset. It was performed on a daily basis during the STW18. Each selected news article is associated with a single transfer move, and it might report a rumour or a verified information. We restricted our harvesting to news containing headlines with clear transfer reports about a single target, even if they report secondary moves in the body text (e.g. another player's alleged move, or interest manifested by other clubs). News with headlines pointing potential transfer of a player to multiple clubs were discarded.

\begin{figure}
    \centering
    \includegraphics[width=0.9\columnwidth]{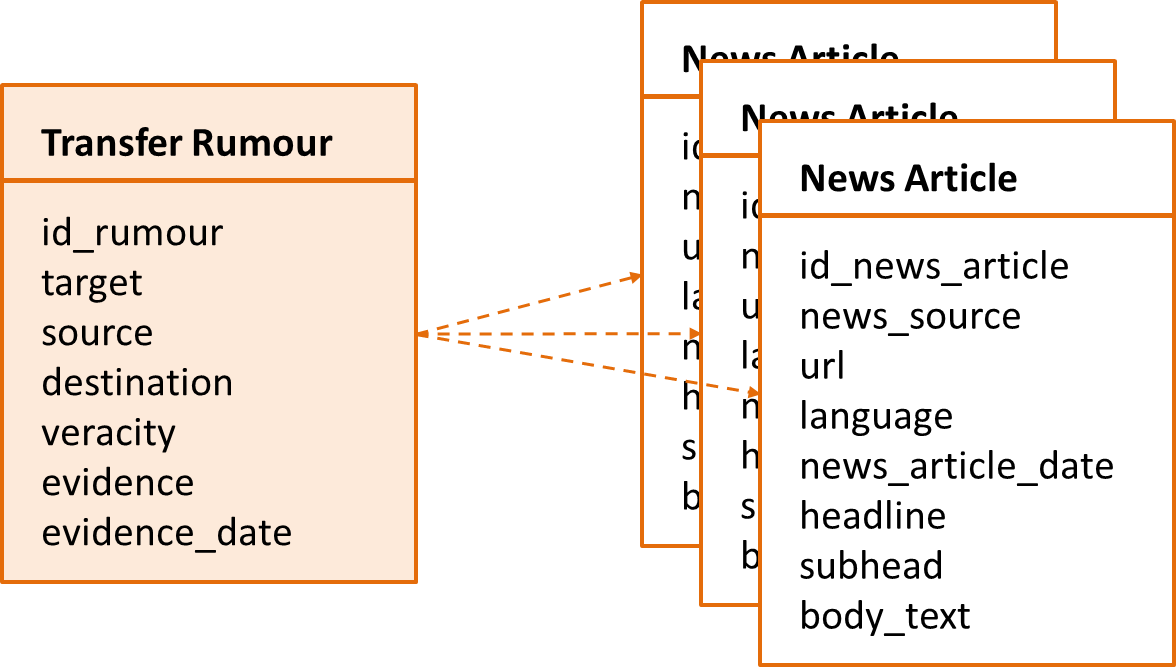}
    \caption{Transfer rumour and news articles attributes.}
    \label{fig:rumour_attr}
\end{figure}

For the news collection, we extracted content and meta-data from each article. Content information includes \textit{headline}, \textit{subhead}, and \textit{body text}, while for the meta-data we registered the \textit{article source}, \textit{language}, \textit{url} and the \textit{publication date} (See Figure \ref{fig:rumour_attr}). The news harvesting task is semi-automated, as we could not build scrapers suited for all news sources due to their poor CSS structure.

\subsection{Rumour reactions crawler}

Once a rumour is identified, we build a crawler using the Twitter's streaming API\footnote{https://developer.twitter.com/en/docs/tweets/filter-realtime/overview.html} to collect posts related to the rumour. In compliance to the established language  restrictions, we only harvest tweets written in the languages of the FTR-18 dataset. We record both the \textit{message content}, \textit{tweet creation date} and the data associated to the user who posted the message (\textit{id}, \textit{screen name}, \textit{description}, \textit{location}, \textit{friends} and \textit{followers count}, \textit{verification account status}). If the post is a retweet, we also register the \textit{retweet status}, the data associated to the user who posted the original message and the \textit{retweet} and \textit{favourite counts}.

Besides collecting tweets related to rumours, we also track Twitter account meta-data and followers statistics either for the players, for the monitored clubs and for the clubs involved in the transfer rumours.

\subsection{Rumour veracity annotation}

As the rumour unfolds, we annotate the rumour veracity by adding to the transfer rumour meta-data the evidences that support or refute the rumour and the publication date of these evidences. Rumour veracity is to be inferred until the end of 2018 Summer Transfer Window. Thus, for the case when source $\neq$ destination, the rumour veracity is stated as \textit{True}, if the transfer is confirmed, or is assigned as \textit{False}, if the target remains in the source club or if the target is transferred to a different destination during STW18. Similarly, for rumours in which source $=$ destination, the rumour veracity is inferred as \textit{False} if the player is transferred, or \textit{True} otherwise.

Despite the challenges of manually determining the credibility of a rumour \cite{zubiaga2018detection}, UEFA registration after STW18 will provide ground truth data for the annotation of the rumours' veracity.

\begin{comment}

\todo[inline]{as bases são as fontes de rumour e depois vai haver informação do encadeamento
gerar conjunto de metadata em volta do rumor}

\end{comment}

\section{Initial Analysis}

\begin{table*}[hbt]
  \caption{Datasets comparison (RD = rumour detection, RT = rumour tracking, SC = stance classification, VC = veracity classification, ER = evidence retrieval).}
  \label{tab:datsets}
\begin{tabular}{ccccccccc}
\toprule
\multirow{2}{*}{Dataset} & \multirow{2}{*}{Data source} & \multirow{2}{*}{\begin{tabular}[c]{@{}c@{}}Publicly\\ Available\end{tabular}} & \multirow{2}{*}{\begin{tabular}[c]{@{}c@{}}Multi-\\ lingual\end{tabular}} & \multicolumn{5}{c}{Usage} \\
 &  &  &  & RD & RT & SC & VC & ER \\
\midrule
Qazvinian's \cite{qazvinian2011rumor} & Social media & \xmark & \xmark & & \cmark   & \cmark &    &    \\
Emergent \cite{silverman2015lies,ferreira2016emergent} & \begin{tabular}[c]{@{}c@{}}News \& Rumour sites\\ \& Social media\end{tabular} & \cmark & \xmark & & & \cmark & \cmark & \\
PHEME-RNR \cite{zubiaga2016learning} & Social media & \cmark & \xmark & \cmark & & & \\
PHEME-RSD \cite{zubiaga2016analysing} & Social media & \cmark & \cmark & \cmark & & \cmark & \cmark & \\
FEVER \cite{thorne2018fever} & Wikipedia & \cmark & \xmark & & & & \cmark  & \cmark \\
LIAR \cite{wang2017liar} & Rumour sites & \cmark & \xmark & & & \cmark & \cmark & \\
FTR-18 & News sites \& Social media & \cmark & \cmark & \cmark & \cmark & \cmark & \cmark &  \\
\bottomrule
\end{tabular}
\end{table*}

Currently, the FTR-18 dataset comprises 3,045 news articles and more than 2,064K tweets (See Table \ref{tab:dataset_stats}). The news articles subset considers transfer news covered by online sports media written in English (1,517), Spanish (747) or Portuguese (781) by 96 different news organisations. This collection includes 304 transfer moves associated with 175 target football players. The Twitter subset contains original messages and retweets written in English (1,130K), Spanish (677K) or Portuguese (257K). The Twitter posts are related to 112 claimed transfer moves involving 84 different football players. The FTR-18 dataset meta-data and corresponding news source scrapers are available at \url{https://github.com/dcaled/FTR-18}.

\begin{table}[h]
  \caption{Current status of FTR-18 dataset.}
  \label{tab:dataset_stats}
    \begin{tabular}{lcc}
    \toprule
    & News articles & Tweets \\
    \midrule
    \textbf{English} & 1,517 & 1,130K \\
    \textbf{Spanish} & 747 & 677K \\
    \textbf{Portuguese} & 781 & 257K \\
    \textbf{Total} & 3,045 & 2,064K \\
    \bottomrule
    \end{tabular}    
\end{table}

In a preliminary analysis of the collected news articles, we noticed many cross-references between different news sources. This pattern has already been identified by \citet{silverman2015lies} and \citet{maia2016jornalismo}. We observed a constant appearance of attribution formulations like ``\textit{source S reported}'' or ``\textit{according to source S}'', followed or not by the link to the referred news source. Other structures used to conceal the origin of the rumour are ``\textit{according to reports/sources}'', ``\textit{are said}'', ``\textit{player P linked to club C}'', ``\textit{reports suggest}''. Expressions like these appear in most of our collected news articles. Another common strategy is reporting an unverified information using news headline as a question \cite{silverman2015lies} (``\textit{Ronaldo out, Neymar in at Real Madrid?}\footnote{https://www.theguardian.com/football/2018/jul/03/football-transfer-rumours-cristiano-ronaldo-real-madrid-juventus-neymar}''). Although less common, this last example is often adopted, mainly by Spanish media.

Another interesting observation concerning transfer news is the cascading move chain associated to a rumour. For example, news sources condition the acquisition of a new player on the sale of another player by the same club (``\textit{Cristiano Ronaldo open to Juventus move as Real Madrid consider selling to fund move for Kylian Mbappe or Neymar}\footnote{https://www.independent.co.uk/sport/football/transfers/cristiano-ronaldo-transfer-news-latest-juventus-real-madrid-neymar-kylian-mbappe-unveil-move-a8434171.html}''). This pattern drives the reader to expect a continuation of the rumourous story and its unfoldings, thus keeping audience's attention for follow-up stories.

\section{Discussion}

FTR-18 is suited for most of the steps involved in a rumour classification process, as presented by \citet{zubiaga2018detection}. Table \ref{tab:datsets} displays a comparison of FTR-18 against existing datasets, showing used data sources, availability, multilingual characteristics and possible usage. At this first phase of the dataset development, we perform annotations on rumours veracity. This task can be finished by the official closing date of STW18 (August 31, 2018), when all football transfers must be concluded. Any subsequent transfers are frozen until the next Winter Transfer Window (January 2019). Once each rumour is labelled on its truthfulness, rumour veracity assessment could be performed on the FTR-18 dataset.

In a second annotation phase, we will manually label our news articles subset, identifying which articles report rumours and which report verified information. This annotation will make FTR-18 suited for a rumour detection task. Stance detection is another possible application for the FTR-18 dataset. The collected content allows the classification of how the news headlines orient towards a given transfer rumour. Besides news' stances, we can evaluate the attitudes manifested by the audience with respect to a transfer move using reactions to Twitter posts. Both news articles labelling and stance annotation depend on human judgement.

The collected data also encourages analysis of rumour tracking. This research topic is scarcely explored and consists of identifying content associated with a rumour that is currently monitored \cite{zubiaga2018detection}. We intend to label the harvested news and Twitter posts a posteriori as related or unrelated to the rumour, following Qazvinian et al's approach \cite{qazvinian2011rumor}. As result, FTR-18 could be used as a rumour dataset suited for binary classification of posts in relation to a rumour.

Finally, the FTR-18 dataset will also allow the analysis on how football transfer news are reported by the sports media. The news articles collection will serve as input for identifying the linguistics structures employed by journalists in transfer coverage. We expect to conduct an investigative work on the propagation patterns present in football transfer news, such as the recurrent conditional transfer moves and the echo effect caused by repetitions of unverified stories published by third-party news organisations. Further improvements in the FTR-18 dataset include the addition of new football leagues (e.g. CONMEBOL), the expansion of monitored clubs set and the collection of data from future transfer windows.

\begin{acks}
This work was supported by FCT, under grant No. UID/CEC/50021/ 2013.
\end{acks}

\bibliographystyle{ACM-Reference-Format}
\bibliography{sample-bibliography}

\end{document}